\documentclass[12pt]{article}
\usepackage{amsmath}
\usepackage{amsfonts}
\def\({\left(}
\def\){\right)}
\def\[{\left[}
\def\]{\right]}
\def\non{ \nonumber }
\begin{document}
\rightline{LPTHE-01-38}
\vskip 2cm
\centerline{\bf Separation of variables for quantum integrable}
\centerline{\bf models related to $ U_q(\widehat{sl}_N) $ }
\vskip 1cm
\centerline {Feodor A. Smirnov}
\vskip 0.5cm
\centerline
{LPTHE, Tour 16, 1-er \'etage, 4, pl. Jussieu,}
\centerline {75252, Paris Cedex 05, France}
\vskip 1cm
{\bf Abstract.} In this paper we construct
separated variable for quantum
integrable models related to the algebra $ U_q(\widehat{sl}_N) $.
This generalizes the results by Sklyanin for $N=2,3$.
\vskip 2cm
\section{Introduction.}
In the papers
\cite{smna}, \cite{sm} we have investigated the relation
between classical and quantum integrable
models and affine Jacobi variety (deformed in quantum case)
for hyper-elliptic curves.
Two works are of fundamental importance for our
investigation. One of them is the book by Mumford \cite{mum},
the other is the work by Sklyanin \cite{skl} who originated the method
of separation of variables in the quantum case.

In the paper \cite{smna} it has been explained that
topological properties of the affine Jacobian are
important for applications to integrable models.
The hyper-elliptic case is very special from this
point of view. It is conjectured in \cite{smna} and
proved in \cite{na} that the cohomologies are extremely
simple in this case. It is clear from the calculation
of Euler characteristic \cite{smna2} that the situation
is much more complicated in the case of more general
spectral curve.
This difference explains our interest in the case of
integrable models related to $sl _N$ for $N\ge 3$.
Let us explain it in some details.

Consider the classical case.
The fundamental object of the theory of
integrable models is the L-operator
which is $N\times N$ matrix
depending on the spectral parameter $x$.
We assume that the L-operator is a polynomial of degree $n$
of special form as described below.
An integrable model of this type is closely
related to affine Jacobian of the spectral curve.
If some conditions of non-degeneracy are satisfied the
genus of this curve equals:
$$g=\frac 1 2 (N-1)(Nn-2)$$

In classics one can construct the separated variables
$(w_j,z_j)$ ($j=1,\cdots ,g$). Relevant for our
applications description of separation of variables in
the classical case is given in the paper \cite{zeit}.
Every pair  $(w_j,z_j)$ is
canonically conjugated. It is equally important that every
pair belongs to the spectral curve. Thus the separated
variables describe a divisor on the spectral
curve which is mapped into affine Jacobian by Abel transformation.

In the quantum case we do not know
an independent definition of deformed Jacobian.
However, following Sklyanin \cite{skl} we are able to
construct for $N=2$ and, which is far more complicated,
for $N=3$ \cite{skl2} the separated variables. This provides the
way of constructing the quantum version of the algebra-geometric
methods.

So, constructing the separated variables is only a part
of our program. But for the case of arbitrary $N$
even this part happens to be highly non-trivial.
That is why we decided to publish it separately.

\section{Separation of variables.}
In this paper we consider standard in R-matrix commutation relations,
but we shall write them in unusual form. Consider the algebra
$U_q(\widehat{sl}_{N})$.  Consider
the evaluation module
related to $N$-dimensional irreducible
module $V$ of the finite-dimensional
subalgebra $U_q(\text{sl}_{N})$:
$$V(x)=V\otimes \mathbb{C}(x)$$
We shall call $V\otimes 1$ the finite-dimensional
subspace of $V(x)$.
Suppose we have some "quantum space" $H$.
Consider  the
L-operator $L_{V(x)}\in \text{End}(H)\otimes\text{End}(V(x))$
As it has been said in the Introduction we shall consider
$L_{V(x)} $ as polynomial of $x$ of degree $n$.
More precisely:
$$L_{V(x)}= L^+(x)+ L^ 0 (x)+ xL^- (x)$$
where $L^+(x)$, $ L^ 0 (x)$ and $ L^- (x)$ are
respectively upper-triangular, diagonal and lower-triangular.
They are respectively polynomials of degree $n-1$, $n$, $n-1$.

The module $V(x)$ should be viewed as an "index".
As operators in $H$ elements of $L_{V(x)} $ satisfy the commutation
relations:
 \begin{align} &R _{V(x),V(y)}
L _{V(x)}L_{V(y)}=
L_{V(y)}L _{V(x)}R _{V(x),V(y)}
\label{RTT}
\end{align}
where the
$U_q(\widehat{\text{sl}}_{N})$ $R$-matrix for evaluation modules
corresponding to vector representations is:
\begin{align}
&R _{V(x),V(y)}=xR_{12}(q)-yR_{21}(q)^{-1}
\label{rmat}
\end{align}
where
$$R_{12}(q)=\sum\limits _{j=1}^N q^{E^{jj}}\otimes q^{E^{jj}}
+(q-q^{-1})\sum\limits _{j>i}E^{ji}\otimes E^{ij},$$
$E^{ij}$ is the matrix with $1$ at the intersection
of $i$-th row and $j$-th column and zeros elsewhere.
We present $L_{V(x)} $ in the form:
\begin{align}
      &L_{V(x)}=\begin{pmatrix} a_{v(x)},b _{v(x)} \non\\
                             c_{v (x)}, d_{v(x)}
 \end{pmatrix}
\end{align}
where as finite-dimensional space
$v(x)$ is $(N-1)$-dimensional
subspace of $V(x)$;
$a_{v(x)}$, $b _{v(x)}$, $ c_{v (x)}$, $d_{v(x)} $ are
respectively scalar, covector, vector and matrix.
We shall use
the operators $b _{v(x)}$ and $d_{v(y)} $, their commutation relations
can be summarized as follows:  \begin{align} &(xq-yq^{-1})b _{v(x)}b _{v(y)}=
b _{v(y)}b _{v(x)}r_{v(x),v(y)}\label{comm} \\
&r_{v(x),v(y)}d _{v(x)}d _{v(y)}= d _{v(y)  }
d_{v(x)}r_{v(x),v(y)} \non\\
&(x-y)b _{v(x)}d _{v(y)} +y(q-q^{-1})s_{v(x),v(y)}d
_{v(x)}b _{v(y)} =
d _{v(y)}b _{v(x)} r_{v(x),v(y)}
\non \end{align}
where $r_{v(x),v(y)}$ is $U_q(\widehat{sl}_{N-1})$
$R$-matrix for vector representations (see (\ref{rmat}),
$$s_{v(x),v(y)}=\sum\limits
_{j=1}^{N-1}e_j^*\otimes e_j,  $$
$e_j^*$ and $e_j$ are dual bases in the finite-dimensional subspaces
of $v^*(x)$ and $v (y)$ respectively.
In order to illustrate our notations let us rewrite the
last equation from (\ref{comm}) in components. 
Denote the components of the covector $b _{v(x)}$ by $b_i(x)$
and the matrix elements of $d _{v(y)}$, $r_{v(x),v(y)} $ by
respectively $d_j^k(y)$, $r_{ij}^{lm}(x,y)$. Then the equation in
question reads:
\begin{align}
&(x-y)b_i(x)d_j^k(y)+y(q-q^{-1})d_i^k(x)b_j(y)
=d_m^k(y)b_l(x)r_{ij}^{lm}(x,y)\non
\end{align}
where the summation over repeated indices is implied. 

In this paper we shall need  evaluation
modules of
$U_q(\widehat{sl}_{N-1})$
which correspond to $q$-deformation of fundamental
representations. Namely, consider the tensor product of $k$
modules ($1\le k \le N-1 $):
$$
v(xq^{-2(k-1)})\otimes \cdots\otimes v(xq^{-2})
\otimes v(x)$$
This module contains an irreducible submodule $w^k(x)$
which can be extracted by applying the projector $p^k$
($q$-antisymmetrizer).
The latter is defined recurrently:
$$p^k=\(x^{-(k-1)}r_{v(xq^{-2(k-1)}),v(xq^{-2(k-2)})} \cdots
r_{v(xq^{-2(k-1)}),v(x)}\)p^{k-1}$$
Notice that $w^{1}(x)=v(x)$, $w^{(N-1)}(x)\simeq \mathbb{C}$.

The relations (\ref{comm}) are covariant in the
following sense. Consider an evaluation
module $w(z)$ and construct the operators:
\begin{align}
&b_{v(x),H\otimes w(z)}=b_{v(x)}r_{v(x),w(z)}\non\\
&d_{v(x),H\otimes w(z)}=d_{v(x)}r_{v(x),w(z)}\non
\end{align}
where we use the notation $r_{v(x),w(z)} $ for the R-matrix
evaluated on corresponding modules.
It is clear that  $b_{v(x),H\otimes w(z)}$
and $  d_{v(x),H\otimes w(z)} $ satisfy the relations (\ref{comm}).

The relations (\ref{comm})  imply a complicated system of commutation
relations of the form:
\begin{align}
&\sum\limits _{q=0}^lX^q_{kl}(x,y)\ b_{v(y),H\otimes v(x)}
(d_{v(y),H\otimes v(x)})^q=\non\\
&=\[\sum\limits _{p=0}^kY^p_{kl}(x,y)\ b_{v(x),H\otimes v(y)}
(d_{v(x),H\otimes v(y)} )^p \]
\frac {r_{v(y),v(x)}}{xq-yq^{-1}}
\label{xb=yb}
\end{align}
The operators $X^q_{kl}(x,y)$ and $Y^p_{kl}(x,y)$ are rather
complicated and we shall need only partial information about
them. First, the leading coefficients are:
\begin{align}
& X^l_{kl}(x,y)=\kappa (x,y)^k
\ b_{v(x)}(d_{v(x)})^k\non\\
&Y^k_{kl}(x,y)=\kappa (x,y)^l
\ b_{v(y)}(d_{v(y)})^l\non
\end{align}
where
$$
\kappa (x,y)=\(\frac{(xq^2-y)(xq^{-2}-y)}{x-y}\)
$$
The rest of these operators can be found from
recurrence relations which follow from (\ref{comm})
using induction. For example
\begin{align}
&X^{q+1}_{k+1,l+1}(x,y)=\kappa (x,y)
\bigl[ X^q_{kl}(x,y)+\non\\&
+\textstyle{\(\frac{x(q - q^{-1})}{x-y}\)}
\sum\limits  _{j=1}
X^{q+j}_{kl}(x,y)\kappa (x,y)^{j-1}
s_{v(y)v(x)}(d_{v(y)})^j s_{v(x)v(y)}\bigr] \ d_{v(x)}
\non\end{align}
\newline
{\bf Lemma 1.} Suppose $l\ge k$, then
\begin{align}
&b_{v(x)}(d_{v(x)})^k\ b_{v(xq^{-2}),H\otimes v(x)}
(d_{v(xq^{-2}),H\otimes v(x)})^l=
\non\\
&=\sum\limits _{p=l-k}^{l-1}\widetilde{X}^{p-l+k}_{kl}(x,xq^{-2})
\ b_{v(xq^{-2}),H\otimes v(x)}
(d_{v(xq^{-2}),H\otimes v(x)})^{p } \label{L1,1}
\end{align}
In particular
\begin{align}
b_{v(x)}\ b_{v(xq^{-2}),H\otimes v(x)} =0
\label{L1,2}
\end{align}
In these formulae $v(x)$ and $ v(xq^{-2})$ are considered as unrelated
modules, i.e. the latter being understood as
$$v(xq^{-2})=v(y)|_{y=xq^{-2}} $$
\newline
{\it Proof.}
Obviously it is sufficient to consider the case $k=l$.
Look at the equation
(\ref{xb=yb}) .
From the recurrence relations cited above one finds
that for $q\ge 0$
$$
X_{kk}^q(x,y)=\kappa (x,y)^k\[\widetilde{X}_{kk}^q(x,y) +O(xq^2-y)\]
$$
where $\widetilde{X}_{kk}^q(x,y)$ is the only term
which does not vanish for $y=xq^{-2}$. It can be shown that
$\kappa (x,y)^{-k}Y_{kk}^p(x,y)$ is finite for $y=xq^{-2}$. The
operator $d_{v(x),H\otimes v(y)}$ contains $r_{v(x),v(y)}$ as right
multiplier. So, the RHS of (\ref{xb=yb}) contains
$$\frac 1 {xq-yq^{-1}} r_{v(x),v(y)}r_{v(y),v(x)}=(xq^{-1}-yq) $$
Dividing  (\ref{xb=yb}) by $\kappa (x,y)^k $ and putting
$y=xq^{-2}$ one proves the Lemma.
\newline
{\bf QED}

\noindent
Now we are ready to introduce the most important definition.

\noindent
{\bf Definition 1.}
The operator $b_{w^k(x), H}\in (w^k(x))^*\otimes \text{End}(H)$
is defined recurrently; $b_{v(x), H}
\equiv b_{v(x)}$ is familiar, further define:
\begin{align}
&b_{w^k(x), H} =x^{-\frac 1 2 k(k-1)}
b_{v(x), H}(d_{v(x), H})^{k-1}\ b_{w^{k-1}(xq^{-2}),
H\otimes v(x) } \label{defb} \end{align} 
The notation $b_{w^{k-1}(xq^{-2}),
H\otimes v(x) }$ stands for 
$b_{w^{k-1}(xq^{-2}),
H}$ in which 
all the $b_{v(xq^{-2l})}$ and 
$d_{v(xq^{-2l})}$ are replaced by
$b_{v(xq^{-2l})}r_{v(xq^{-2l}),v(x)}$ and 
$d_{v(xq^{-2l})}r_{v(xq^{-2l}),v(x)}$.  

It is easy to observe that
the RHS of (\ref{defb}) contains effectively the projector $p^k$
acting from the right, hence it belongs to $(w^k(x))^*$, so,  the
definition is consistent.  Lemma 1 has important
\newline {\bf
Corollary.} The following vanishing property holds:
\begin{align}
&b_{v(x), H}(d_{v(x), H})^k\ b_{w^{l}(xq^{-2}), H\otimes v(x) }=0
\non\\ &\text{for}\qquad k<l \label{van}
\end{align}
{\it Proof.}
Write down the  explicit expression for  $b_{w^l(x), H}$ using $l-1$
times (\ref{defb}).  Then apply repeatedly the relation (\ref{L1,1})
lowering the degree $k$. At some stage the degree will turn to zero,
then the relation (\ref{L1,2})  is applicable.  \newline {\bf QED}

It has been mentioned that $w^{(N-1)}(x)\simeq \mathbb{C}$
which means that
\begin{align}
&b_{w^k(x), H\otimes w^{(N-1)}(y)}=\varphi _k(x,y) \ b_{w^k(x), H}
\label{N-1=0,1}
\end{align}
where $\varphi _k(x,y)$ is a  $\mathbb{C}$-number function.
The latter can be calculated explicitly using known formula
for quantum determinant of R-matrix:
$$\varphi _k(x,y) =\prod\limits _{j=0}^{k-1}\((xq^{-(2j+1)}-yq^{1-2N})
\prod\limits _{i=0}^{N-3}(xq^{-2j}-yq^{-2i})\)^{k-j}$$
The
most important property of the operators $b_{w^k(x), H} $
is given by
\newline
{\bf Lemma 2.} The relation holds
\begin{align}
&\chi _{k,l}(x,y)
b_{w^{k}(x), H}\ b_{w^{l}(y), H\otimes w^{k}(x) } =\non \\
&=\chi _{l,k}(y,x)
b_{w^{l}(y), H}\ b_{w^{k}(x), H\otimes w^{l}(y) } \ \psi _{l,k}(y,x)
\ r_{w^{l}(y), w^{k}(x)}
\label{comb}
\end{align}
where the $\mathbb{C}$-number functions are given by
\begin{align}
& \chi _{k,l}(x,y)=\prod\limits _{j=0}^{k-1}\(
\prod\limits _{i=0}^{l-1}\kappa (xq^{-2j},yq^{-2i})\)^{k-j} ,
\non\\
&\psi _{l,k}(y,x) =\prod\limits _{j=0}^{k-1}\prod\limits _{i=0}^{l-1}
\(yq^{-2i-1}-xq^{-2j+1} \)^{-1}\non
\end{align}
{\it Proof}. Rewrite the commutation relation (\ref{xb=yb}) in the
form:
\begin{align}
&\kappa(x,y)^i \ b_{v(x),H} (d_{v(x),H} )^i
\ b_{v(y),H\otimes v(x)} (d_{v(y),H\otimes v(x)})^j
=\non\\ &=
\kappa (y,x)^j\[
b_{v(y),H} (d_{v(y),H})^j
\ b_{v(x),H\otimes v(y)} (d_{v(x),H\otimes v(y)} )^i
\] \frac {r_{v(y),v(x)}}{xq-yq^{-1}}  +
\non\\
&+\text{"unwanted terms"} \label{h1}
\end{align}
where
\begin{align}
&\text{"unwanted terms"} =\non\\
&-\sum\limits _{q=0}^{j-1}X^q_{ij}(x,y)\ b_{v(y),H\otimes
v(x)} (d_{v(y),H\otimes v(x)})^q+\non\\ &+\[\sum\limits
_{p=0}^{i-1}Y^p_{ij}(x,y)\ b_{v(x),H\otimes v(y)}
(d_{v(x),H\otimes v(y)}
)^p \] \frac {r_{v(y),v(x)}}{xq-yq^{-1}} \non\end{align}
We prove the relation (\ref{comb}) using induction with
respect to $k$ and $l$. The definition (\ref{defb}) implies that
we have to use the relations  (\ref{h1}) in the inductive
procedure. The most important point of the proof is that
due to the relations (\ref{van}) the contributions
from "unwanted terms" vanish.
\newline
{\bf QED}

In what follows we shall need the formula:
\begin{align}
&r_{w^{l}(y), w^{N-1}(x)} = \rho_l(y,x)I,\non\\
&\rho_l(y,x)=\prod\limits _{i=0}^{l-1}
(yq^{-2i+1}-xq^{3-2N})\prod\limits _{j=0}^{N-3}
(yq^{-2i}-xq^{-2j})
\label{N-1=0,2}
\end{align}
Now we are in position to define the basic objects
\newline
{\bf Definition 2.} $B(x)=b_{w^{N-1}(x),H}$.
\vskip 0.8cm
\noindent
{\bf Theorem 1.} The operators $B(x)$ generate a commuting family:
\begin{align}
&[B(x),B(y)]=0\label{[BB]}
\end{align}
{\it Proof.} Due to relations (\ref{comb}) and
(\ref{N-1=0,1}) and (\ref{N-1=0,2} ) the proof
reduces to the identity for $\mathbb{C}$-number functions:
$$\chi _{N-1,N-1}(x,y)\varphi _{N-1}(y,x)=$$
$$=\chi _{N-1,N-1}(y,x)  \varphi _{N-1}(x,y)
\psi _{N-1,N-1}(y,x) \rho _{N-1}(y,x)$$
This identity is proved by direct calculation.
\newline
{\bf QED}

The operator $B(x)$ is a polynomial of degree
$g=\frac 1 2 (N-1)(Nn-2)$. Due to Theorem 1
it can be developed as
$$ B(x)=\beta \prod\limits _{j=1}^g(x-z_j)$$
where all operators $\beta, z_1,\cdots , z_g$
mutually commute. The operator $\beta $ is a kind of
zero-mode, it is not very interesting for us.
We want to define the operators canonically conjugated
to $z_1,\cdots , z_g$.
To this end we have to define new object.
\newline
{\bf Definition 3.} Consider some covector $\xi\in
(\mathbb{C}^{N-1})^*$.  By $\xi _{v(x)}$ we denote $\xi$ in
the finite-dimensional subspace  of $v ^*(x)$. Define
\begin{align}
&Y(x)=\xi _{v(x)} b_{w^{N-2}(xq^{-2}),H\otimes v(x)}, \non\\ &X(x)=\xi
_{v(x)}d_{v(x)}b_{w^{N-2}(xq^{-2}),H\otimes v(x)}\non \end{align} and
$$D(x)=Y^{-1}(x)X(x)$$
The operator $D(x)$ depends upon the  choice of $\xi$, however,
the final objects we are interested in ($w_j$) are independent
of $\xi$ as is explained later.
By already familiar calculations we get
\newline
{\bf Lemma 3.} The relation holds
\begin{align}
&b_{w^k(x),H\otimes v(y)}\ r_{v(y),w^k(x)}\ d_{v(y)}=\non\\
&=\sigma _k(x,y)\ d_{v(y)}\ b_{w^k(x),H\otimes v(y)}  +
U(x,y)b_{v(y)}\label{db}
\end{align}
where  $U(x,y)$ is some irrelevant for our goals operator,
$$\sigma _k(x,y)=\prod\limits _{j=0}^{k-1}\kappa (xq^{-2j},y)$$
The commutation relation between $D$ and $B$ is given by
\newline
{\bf Theorem 2.} The relation holds
\begin{align}
&B(x)D(y)=\frac {xq-yq^{-1}}{x-y}D(y)B(x)  + Z_{N-2}(x,y) B(y)
\label{AB}
\end{align}
the operator $Z_{N-2}(x,y)$ is defined below.
\newline
{\it Proof.} The relation (\ref{comb}) implies that
$$B_{H\otimes v(y)}(x)\ b_{w^{N-2}(yq^{-2}),H\otimes v(y)}
=\mu(x,y)\ b_{w^{N-2}(yq^{-2}),H\otimes v(y)} \ B(x)$$
with some irrelevant for us function $\mu (x,y)$,
the notation $B_{H\otimes v(y)}(x)$ should be clear.
Multiplying this equation by $\xi _{v(y)} $ or by $\xi
_{v(y)}d _{v(y)} $, and using the relation (\ref{db}) one finds two
equations:
\begin{align} &\eta
_{v(y)}(x,y)\ b_{w^{N-2}(yq^{-2}),H\otimes v(y)}=\mu
(x,y)Y(y)B(x),\non\\
&\eta _{v(y)}(x,y)\ d_{v(y)}\ b_{w^{N-2}(yq^{-2}),H\otimes v(y)}=\mu
(x,y)\sigma _{N-1}(x,y)X(y)B(x),\label{t2}
\end{align}
where
$$\eta _{v(y)}(x,y)  =\xi _{v(y)} B_{H\otimes v(y)}(x)$$
Notice that $\eta _{v(y)}(x,y)$ is
a covector in finite-dimensional subspace of $v^*(y)$. Thus we
can develop it with respect to the following basis:
\begin{align}
&\eta _{v(y)}(x,y) =Z(x,y)\ \xi _{v(y)}+\sum\limits _{j=1}^{N-2}
Z_j(x,y)\ b_{v(y)}(d_{v(y)} )^{j-1}
\label{eta}
\end{align}
where $Z(x,y),Z_j(x,y)$ are operator-valued coefficients.
We shall need two of them: $Z(x,y)$ and $ Z_{N-2}(x,y)$.
In order to find the first of them we multiply (\ref{eta})
by $b_{w^{N-2}(yq^{-2}),H\otimes v(y)}$ and use the property
(\ref{van}).  This gives $$Z(x,y)=\(\eta
_{v(y)}(x,y)b_{w^{N-2}(yq^{-2}),H\otimes v(y)}\) Y^{-1}(y)$$
Calculation of $ Z_{N-2}(x,y)$  is somewhat more complicated.
Multiply (\ref{eta}) by $b_{w^{N-3}(yq^{-2}),H\otimes v(y)}$.  The
result is a covector from $w^{N-2}(y) $. The finite-dimensional part
of the latter space is isomorphic to $\mathbb{C}^{N-2}$.  Consider a
vector $\nu $ from this finite dimensional part and convolute the
equation  (\ref{eta})    with $\nu $.  Then using already known
expression  for $Z(x,y)$ one finds:
\begin{align} & Z_{N-2}(x,y)
=\eta _{v(y)}(x,y)\bigl(1-b_{w^{N-2}(yq^{-2}),H\otimes v(y)}
Y^{-1}(y)\xi _{v(y)}\bigr)
\non\\
&\times
b_{w^{N-3}(yq^{-2}),H\otimes v(y)}  \ \nu _{w^{N-2}(y)}
\widetilde{Y}^{-1}(y)\non \end{align}
where
$$\widetilde{Y}(y) =b_{w^{N-2}(y),H}   \ \nu  _{w^{N-2}(y)}$$

Substitute (\ref{eta})
into (\ref{t2}) and use (\ref{van}):  \begin{align}
&Z(x,y)Y(y)=\mu
(x,y)Y(y)B(x),\non\\
&\rho _1(y,x) \bigl(Z(x,y)X(y)+Z_{N-2}(x,y)B(y)\bigr)=
\non\\
=&\mu
(x,y)\sigma _{N-1}(x,y)X(y)B(x)
\end{align}
From these equations the relation (\ref{AB}) follows immediately.
\newline {\bf QED}

Following Sklyanin \cite{skl2} we define the operators
$$w_j=\int\limits _CD(y)\frac {dy}{y-z_j}$$
where $C$ encircles $z_j$. The meaning of this definition is clear:
we substitute $z_j$ as argument of $D(y)$ from the right.
Apply this operation to the equation  (\ref{AB}).
The term $Z_{N-2}(y)B(y)$ vanishes because of presence on $B(y)$.
Certainly, one should be careful because $Z_{N-2}(x,y)$
contains $Y(y)$ and $\widetilde{Y}(y)$ in denominator, but
these operators cannot cancel $B(y)$.
Thus
$$w_iz_j=q^{2\delta _{ij}}z_jw_i, \qquad w_j\beta=q^{-1}\beta w_j  $$

Let us show that $w _j$ does not depend on the choice of $\xi$.
Take another covector $\xi '$ and write the formula:
$$\xi ' _{v(x)}=U(x)\xi _{v(x)}+\sum\limits _{j=1}^{N-2}
U_j(x)b_{v(x)} (d_{v(x)})^{j-1}$$
with some $U$, $U_j$ which can be found explicitly.
From this formula one easily finds that
$$ D_{\xi '}(x)=D_{\xi }(x) +Y_{\xi '}(x)^{-1}U_{N-2}(x)B(x)$$
where dependence on $\xi$ and $\xi '$ is marked explicitly.
The last term vanishes when $z_j$ is substituted from the right.
So, $w_j$ does not depend on $\xi$.

It remains to prove that
\begin{align}
[w_i,w_j]=0
\label{ww}
\end{align}
To this end we shall prove the
\newline
{\bf Theorem 3.} The following commutation relations hold:
\begin{align}
&D(x)D(y)=D(y)D(x)+S(x,y)B(x)+ T(x,y)B (y)
\label{AA}
\end{align}
where $S(x,y)$, $T(x,y)$ are some operators which can be found
explicitly.
\newline{\it Proof.} Consider the relation \begin{align} &
b_{w^{N-2}(xq^{-2}),H\otimes v(y)\otimes v(x)}
\ b_{w^{N-2}(yq^{-2}),H\otimes v(y)\otimes v(x)\otimes
w^{N-2}(xq^{-2})}=\non\\
&=\mu (x,y)b_{w^{N-2}(yq^{-2}),H\otimes v(y)\otimes v(x)}
\ b_{w^{N-2}(xq^{-2}),H\otimes v(y)\otimes v(x)
\otimes
w^{N-2}(yq^{-2})}
\non\\ &\times r_{w^{N-2}(yq^{-2}),w^{N-2}(xq^{-2})  }\label{t31}
\end{align}
where $ \mu (x,y)  $ is some irrelevant function.
The operators $b_{w^{N-2}(xq^{-2}),H\otimes v(y)\otimes v(x)}$,
$b_{w^{N-2}(yq^{-2}),H\otimes v(x)\otimes v(y)}$ contain
effectively projectors on the subspaces respectively
$w^{N-1}(x) $ and $w^{N-1}(y) $. Hence  (\ref{t31}) can be rewritten
as
\begin{align}
&
b_{w^{N-2}(xq^{-2}),H\otimes v(y)\otimes v(x)}
\ b_{w^{N-2}(yq^{-2}),H\otimes v(y)}=\non\\
&=\widetilde{\mu} (x,y)b_{w^{N-2}(yq^{-2}),H\otimes v(y)\otimes v(x)}
\ r_{v(y),w^{N-2}(yq^{-2})}
\ b_{w^{N-2}(xq^{-2}),H\otimes v(x)}
\label{t32} \end{align}
Multiply (\ref{t32}) by $\xi _{v(x)}$, introduce the notation
$$\eta _{v(x),H\otimes v(y)}(x,y)=
\xi _{v(x)}\ b_{w^{N-2}(yq^{-2}),H\otimes v(y)\otimes v(x)}
\ r_{v(y),w^{N-2}(yq^{-2})}
$$
and use the same trick as in the proof of Theorem 2:
$$\eta _{v(x),H\otimes v(y)} =
V(x,y)_{H\otimes v(y)}\xi _{v(x)} +\sum\limits _{j=1}^{N-2}
V_j(x,y)_{H\otimes v(y)}
\ b_{v(x)}(d_{v(x)} )^{j-1}$$
the explicit formulae for $V(x,y)_{H\otimes v(y)}$ and
$V_{N-2}(x,y)_{H\otimes v(y)} $
can be found as it has been done in the proof of
Theorem 2.
Hence
\begin{align}
Y_{H\otimes v(y)}\ b_{w^{N-2}(yq^{-2}),H\otimes v(y)}=
\widetilde{\mu} (x,y)V(x,y)_{H\otimes v(y)}Y(u)
\end{align}
Further, using Lemma 3 one gets
\begin{align}
&X_{H\otimes v(y)}\ b_{w^{N-2}(yq^{-2}),H\otimes v(y)}=
\non\\
&=\sigma _{N-2}(x,y)
\widetilde{\mu} (x,y)\bigl(V(x,y)_{H\otimes v(y)}Y(x)+ V_{N-2}
(x,y)_{H\otimes v(y)}B(u)\bigr)
\non
\end{align}
Combining these two equations one finds
\begin{align}
&b_{w^{N-2}(yq^{-2}),H\otimes v(y)}
D(x)+V_{N-2}(x,y)_{H\otimes v(y)}B(x)
=\non\\&=
\sigma _{N-2}(x,y)
D_{H\otimes v(y)}(x)b_{w^{N-2}(yq^{-2}),H\otimes v(y)}
\non
\end{align}
Let
$$\zeta _{v(y)}(x,y)=\xi _{v(y)}D_{H\otimes v(y)} (x)$$
Then
$$\zeta _{v(y)}(x,y)= W(x,y)\xi _{v(y)}+
\sum\limits _{j=1}^{N-2}W_j(x,y)b_{v(y)}
(d_{v(y)}  )^{j-1}$$
and
\begin{align}
&Y(y)D(x)+\xi _{v(y)}V_{N-2}(x,y)_{H\otimes v(y)}B(u) =
\sigma _{N-2}(x,y)W(x,y)Y(y)
\label{t37}
\end{align}
Using Lemma 3 one shows that
$$d_{v(y)}D_{H\otimes v(y)}(x)= D_{H\otimes v(y)}(x)d_{v(y)}
+ (\text{smth}) b_{v(y)} $$
Using this equation and (\ref{van}) one finds:
\begin{align}
&X(y)D(x)+\xi _{v(y)}V_{N-2}(x,y)_{H\otimes v(y)}B(u) =
\non\\ &=\sigma
_{N-2}(x,y)\bigl (W(x,y)Y(y)+W_{n-2}(x,y)B(v)\bigr)
\label{t38}
\end{align}
From (\ref{t37}), (\ref{t38}) and Theorem 2
one proves 
Theorem 3 follows immediately.
We do not write explicit expressions for $S(x,y)$ and
$T(x,y)$, they contain some denominators which do not
cancel $B(x)$ and $B(y)$.
\newline {\bf QED}
\newline
The commutation relations (\ref{ww}) follow from this theorem.

\section{Conclusion.}
Let us mention two problems which has not been
addressed in this paper.

First, it should be shown that every observable in the
model can be expressed in terms of separated variables.
This is actually not the case. There are some additional
degrees of freedom. The best way out of this difficulty is
to perform a reduction eliminating this additional degrees
of freedom but keeping the reduced algebra  of observables
closed. This procedure has been performed in \cite{sm} for $N=2$
in quantum case and in \cite{zeit} for arbitrary $N$ in classics.
The reduction is possible in the quantum case for arbitrary $N$,
but we would like to postpone the publication of corresponding
bulky calculations.

Second, the fact that the separated variables belong
to the spectral curve is replaced in the quantum
case by Baxter equation . We do not deduce this equation
here, again we shall do it elsewhere.
\vskip 0.2cm
\noindent
{\it Acknowledgment.} This work was partly supported by the INTAS
grant 
\newline INTAS 00-00055 and JSPS fellowship.

\end{document}